\documentstyle[preprint,12pt,aps,tighten,floats,epsf]{revtex}

\def\D0bar{\overline D{}^0}
\def\Zres{Z_{\rm res}}
\def\Zsut{Z_{SU(3)}}

\begin{document}

\preprint{
\vbox{
         \hbox{IASSNS-HEP-99-104}
         \hbox{JHU--TIPAC--99010}
         \hbox{hep-ph/9911369}
         \hbox{}\hbox{} }}

\title{Strong phases and $D^0-\D0bar$ mixing parameters}
\author{Adam F. Falk${}^{a}$, Yosef Nir${}^{bc}$ and
Alexey A. Petrov${}^{a}$}
\address{ \vbox{\vskip 0.truecm}
$(a)$ Department of Physics and Astronomy, The Johns Hopkins
University \\
    Baltimore, Maryland 21218 USA \\
$(b)$ Department of Particle Physics,
          Weizmann Institute of Science \\ Rehovot 76100, Israel\\
$(c)$ School of Natural Sciences, Institute for Advanced Study \\
          Princeton, New Jersey 08540 USA\footnote{Address for
          academic year 1999-2000; nir@sns.ias.edu}}

\maketitle
\thispagestyle{empty}
\setcounter{page}{0}
\begin{abstract}
We argue that there could be significant $SU(3)$ violating resonance
contributions to $D\to K\pi$ decays which would affect the extraction
of the $D^0 - \D0bar$ mixing parameters from experiment.
Such contributions can induce a strong phase in the interference
between the doubly Cabibbo suppressed and the mixing induced Cabibbo
favored contributions to the $D^0\to K^+\pi^-$ and $\D0bar\to K^-\pi^+$
decays. Consequently, the interpretation of a large,
CP conserving interference term can involve a large mass difference
$\Delta M$ rather than a large width difference~$\Delta\Gamma$.

\bigskip\bigskip
\centerline{(Revised December 1999)}

\end{abstract}

\newpage

%%%%%%%%%%%%%%%%%%%%%%%%%%%%%%%%%%%%%
\section{Introduction}

Due to the smallness of the Standard Model (SM)
$\Delta C = 2$ amplitude, $D^0 - \D0bar$ mixing offers a unique
opportunity to probe flavor-changing interactions which may be
generated by new physics at short
distances~\cite{Datta:1985jx,Petrov:1997ch,%
Donoghue:1986hh,Wolfenstein:1985ft,%
Georgi:1992as,Ohl:1993sr,Golowich:1998pz}. In fact, if $D^0 -
\D0bar$ mixing is observed in the current round of experiments, it
would unambiguously signal new flavor physics beyond the Standard
Model.  In particular, while the $D^0 - \D0bar$ mixing parameters
$\Delta M$ and $\Delta\Gamma$ are small in the Standard Model, $\Delta
M$ could be enhanced significantly by new short-distance
interactions.  An enhancement of $\Delta\Gamma$ is considerably less
likely, since there are already strong constraints on the branching
ratios of $D^0$ and $\D0bar$ to common final states. Experimentally,
$\Delta M$ and $\Delta \Gamma$ are measured by studying the time
evolution of neutral $D$ mesons decaying to a particular final state.
The two body decays $D \to K \pi$ are particularly popular, due to
their relatively simple experimental signature~\cite{Artuso:1999hy}.

The phenomenon of $D^0 - \D0bar$ mixing occurs because the neutral $D$
meson mass eigenstates do not coincide with the flavor eigenstates.
The former are related to the latter by
$|D_{1,2}\rangle=p|D^0\rangle\pm q|\D0bar\rangle$, with
$|p|^2 +|q|^2=1$.  Since it is the mass and not the flavor eigenstates
which evolve diagonally, a state which is produced at $t=0$ as a
$D^0$ meson may be detected at some later time $t$ as a $\D0bar$. For
the method of detecting $D^0 - \D0bar$ mixing involving the $D^0 \to
K\pi$ decay mentioned above, $D^0 - \D0bar$ mixing contributes
through  the sequence $D^0\to\D0bar\to K^+\pi^-$, where the second
stage is Cabibbo favored (CF). The search is complicated by the
presence of a direct, doubly Cabibbo suppressed (DCS) process. Since
the two sequences $D^0\to\D0bar\to K^+\pi^-$ and
$D^0\to K^+\pi^-$ lead to the same final state, one must study the
time-dependent decay rate in order to separate these two
contributions.  These decay rates are given by
\begin{eqnarray} \label{evolution}
\Gamma (D^0(t) \to K^+\pi^-) &=&
\textstyle{\frac{1}{4}} e^{- \Gamma t}\, | B |^2\,
\left | {q}/{p} \right |^2 \times
\nonumber \\
&& \ \left \{
\left( \Delta M^2 + \textstyle{\frac{1}{4}} \Delta \Gamma^2
\right) t^2 + \Big( 2 \mbox{Re} (\lambda) \Delta \Gamma +
4 \mbox{Im} (\lambda) \Delta M \Big)t +
4 | \lambda |^2 \right \},
\nonumber
\\
\Gamma (\D0bar(t) \to K^-\pi^+ ) &=&
\textstyle{\frac{1}{4}} e^{- \Gamma t}\, | \overline B|^2\,
\left | {p}/{q} \right |^2 \times
\nonumber \\
&& \ \left \{
\left( \Delta M^2 + \textstyle{\frac{1}{4}} \Delta \Gamma^2
\right) t^2 + \Big( 2 \mbox{Re} (\overline \lambda) \Delta \Gamma +
4 \mbox{Im} (\overline \lambda) \Delta M \Big)t +
4 | \overline \lambda |^2 \right \},
\end{eqnarray}
where the quadratic terms ($t^2 e^{-\Gamma t}$)
arise from $D^0 -\D0bar$ mixing,
the linear terms ($t e^{-\Gamma t}$)
describe interference between mixing and DCS
contributions, and the constant terms
($e^{-\Gamma t}$) describe purely
DCS decays. The importance of the linear term in
Eq.~(\ref{evolution}) was emphasized in Ref.~\cite{Blaylock:1995ay}.
The mass and width differences are defined as $\Delta M=M_2-M_1$ and
$\Delta\Gamma=\Gamma_2-\Gamma_1$.  We also define
\begin{equation} \label{lambda}
\lambda = |\lambda| e^{i (\delta+\theta)} =
\frac{p}{q}\, \frac{A}{B}\,,\qquad
\overline\lambda = |\overline \lambda| e^{i(\delta-\theta)} =
\frac{q}{p}\, \frac{\overline A}{\overline B}\,,
\end{equation}
to represent the convention-independent ratios of the amplitudes
\begin{eqnarray}
A &=& \langle K^+\pi^- | H | D^0 \rangle\,,\qquad
\overline A = \langle K^-\pi^+ | H | \D0bar \rangle\,,
\nonumber\\
B &=& \langle K^+\pi^- | H | \D0bar \rangle\,,\qquad
\overline B = \langle K^-\pi^+ | H | D^0 \rangle\,.
\end{eqnarray}
The amplitudes $B$ and $\overline B$ are Cabibbo favored, while $A$
and $\overline A$ are doubly Cabibbo suppressed; $\delta$ and
$\theta$ define strong and weak phases, respectively, of $\lambda$
and $\bar \lambda$. In
the limit of unbroken $SU(3)$ symmetry, $A$ and $B$ are simply related
by CKM factors,
$A = (V_{cd} V_{us}^*/V_{cs} V_{ud}^*) B$.  In particular, $A$ and $B$
have the same strong phase~\cite{Wolfenstein:1995kv}, leading to
$\delta=0$ in Eq.~(\ref{lambda}).  Neglecting the small weak phases in
the CKM elements of the first two generations, that is, setting
$\theta\approx\arg(V_{cd}V_{ud}^*/V_{cs}V_{us}^*)=0$, it is often assumed
then that $\lambda$ and $\bar\lambda$ are real.
The linear terms in Eq.~(\ref{evolution}) are therefore dominated by
the terms proportional to $\Delta \Gamma$. More generally, for $\delta=0$
and any $\theta$, same-sign contributions to the interference terms in the
two CP-conjugate processes of Eq.~(\ref{evolution}) are proportional
to $\Delta \Gamma$. From this perspective, the experimental
observation of such anomalously large coefficient of $t$ would be
extremely surprising, since a substantial enhancement of
$\Delta\Gamma$ is difficult to explain even in the presence of new
physics.  It is clearly important to explore the robustness of this
conclusion.

This argument that $\delta$ vanishes, based on $SU(3)$ symmetry, has
the virtue of being model-independent.  However, $SU(3)$ is known
experimentally to be broken badly in some $D$ decays,\footnote{For
example, $\Gamma(D^0 \to K^+ K^-) / \Gamma(D^0 \to \pi^+\pi^-) =
2.75 \pm 0.15 \pm 0.16$ experimentally~\cite{Aitala:1998ff}, while the
ratio is predicted to be one in the $SU(3)$ limit.} and therefore we might
expect the relative strong phase of $A$ and $B$ not to vanish.  The
existing hadronic models which incorporate $SU(3)$ violation seem to
prefer a small value of this phase, $\sin\delta\alt0.2$, with most models
giving
$\sin\delta\alt0.1$~\cite{Chau:1994ec,Browder:1996ay}.

Do the small phases found in these models imply that the CP conserving
part in the linear term is still dominated by $\Delta \Gamma$?  Probably not.
First, it is likely that $\Delta \Gamma$ is quite small, as it is generated by
physical intermediate states and thus is not sensitive to any new
$\Delta C = 2$ interactions that
might enhance $\Delta M$~\cite{Golowich:1998pz}.  In new physics
scenarios of enhanced $D^0-\D0bar$ mixing, $\Delta M\gg\Delta\Gamma$,
and hence any nonzero coefficient of $\Delta M$, even if fairly small,
permits it to dominate the linear term.

Second, these hadronic models may well not have enough $SU(3)$
violation in them, which is crucial since the phase $\delta$ is an
$SU(3)$ violating effect.  In particular, they may have difficulty
accomodating the observed rates for the very mode used for the
$D^0-\D0bar$ mixing studies.  The ratio~\cite{Cinabro:1993nh}
\begin{equation} \label{ratio}
R = \frac{{\cal B}(D^0 \to K^+ \pi^-)}{{\cal B}(\D0bar \to K^+ \pi^-)}
\left |\frac{V_{ud} V_{cs}^*}{V_{us} V_{cd}^*}\right |^2
\end{equation}
is unity in the $SU(3)$ symmetry limit.  However, the world average for
this ratio is~\cite{Artuso:1999hy,Cinabro:1993nh,Aitala:1998fg,%
Barate:1998uy}
\begin{equation}
R_{\rm exp}=1.58\pm0.49\,,
\label{Rexp}
\end{equation}
computed from the individual measurements using the standard methods of
Ref.~\cite{Caso:1998tx}.  It is quite possible that $SU(3)$ is
badly broken in $D\to K \pi$ transitions, in which case a significant
strong phase of $\lambda$ would be a natural consequence.

It has been suggested that large $SU(3)$ violation in $D\to K\pi$
could be accomodated in a factorization approach by a conspiracy of
individually small $SU(3)$ violating effects~\cite{Chau:1994ec}.
However, the use of the factorization anzatz is at least dubious in
$D$ decays. It is also interesting to note that new experimental data
on $D$ decay form factors~\cite{Bartelt:1997vu},
\begin{equation}
\left | {f_+^{D\pi}}/{f_+^{DK}} \right |^2 = 0.9 \pm 0.3 \pm 0.3\,,
\end{equation}
seem to suggest that $f_+^{D\pi}(0) < f_+^{DK}(0)$, while the
analysis of Ref.~\cite{Chau:1994ec} requires the opposite inequality.
In any case, one may doubt our ability to predict accurately even the
magnitude of the ratio of DCS to CF amplitudes.  The phase $\delta$
is presumably still more uncertain.

The purpose of this paper is to make this suggestion concrete, by
displaying a simple model in which $\lambda$ can easily have a significant
strong phase.  The model is based on the hypothesis of
significant contributions to $D^0\to K^+\pi^-$ and $\D0bar\to
K^+\pi^-$ from nearby resonances.  A deviation of $R_{\rm exp}$ from unity
would imply, within our model, that the resonance couplings violate
$SU(3)$ symmetry, in which case a large strong phase difference between
the amplitudes $A$ and $B$ is a generic consequence.  Therefore, it is
unjustifiable to neglect the term proportional to ${\rm
Im}\lambda\cdot\Delta M\cdot te^{-\Gamma t}$ in the time-dependent
decay rate, even in the CP limit.  No prejudice whatsoever about the
phase of $\lambda$ should pollute the experimental analysis.

%%%%%%%%%%%%%%%%%%%%%%%%%%%%%%%%%%%%%%%%%%%%%%%%%%
\section{Resonances as a source for strong phases}

Our goal is to investigate the possible effects of
broken flavor $SU(3)$ on the relative strong phases of the CF and DCS
amplitudes $A$ and $B$.  We will exploit the fact that the
mass of the $D$ meson is {\it small} enough that it lies in the region
populated by light quark resonances~%
\cite{Golowich:1998pz,Golowich:1981yg,Close:1996cf,Gronau:1999zt}.
(In this respect, $D$ decays are very different from $B$ decays, which
lie far above the light resonance region.  See
Ref.~\cite{Petrov:1999iv} for a recent review.) It is therefore quite
conceivable that a strong phase difference is generated by processes
in which these resonances appear as $s$ channel intermediate states.
If there is large $SU(3)$ violation related to these resonance
effects, the phase difference $\delta$ could also be large.

For the sake of simplicity, we will assume that all or most
$SU(3)$ violation in $(D^0,\D0bar)\to K^+\pi^-$ comes from resonance
contributions to the decay.  Rather than modeling the resonance
couplings directly, in our model we will fit them to the value of
$R_{\rm exp}$.  This is clearly an approach designed to maximize $SU(3)$
violation in the resonance contributions, which will typically
maximize the phase difference $\delta$.  Our point is {\it not\/} that
we consider it likely that such a simple model accounts
accurately for the phenomenology of these decays.  Rather, it is,
first, that these resonance contributions are generically present, and
second, that without any tuning of the parameters it is quite {\it
possible\/} to generate nonnegligible $\delta$.  Therefore, we would
argue, the existence of models in which $\delta$ is found to be small
is hardly sufficient for one to conclude that this is not the
case.

For the decays at hand the most important resonances are the ``heavy
kaons'' of positive parity, $K^*(1430)$ and $K^*(1950)$.  Both of these
resonances are quite broad, $\Gamma(K^*(1430))=287 \pm 10 \pm 21$, and
$\Gamma(K^*(1950))=201 \pm 34 \pm 79$~\cite{Caso:1998tx}.  We will
refer to $K^*(1430)$ and $K^*(1950)$ generically as ``$K_H$.''  We
emphasize that the positions of these resonance poles are not very well
established.

We will introduce an effective phenomenological coupling of the light
quark resonance to a $D^0$ or $\D0bar$ meson, which we will fit to the
experimental data.  We note in passing that most models of final
state interactions in $D$ decays employ resonance dominance in some
form (however, see also Ref.~\cite{Gerard:1998un}).  For instance,
Refs.~\cite{Buccella:1995nf,Elaaoud:1999pj} employ a set of two body
intermediate states which rescatter to the final state via a resonance.
Essentially, the coupling of the resonance to the $D$ meson is
modeled by this two body contribution. In
Refs.~\cite{Golowich:1998pz,Petrov:1997fw} a similar coupling was
taken to be dominated by a contact interaction. In our
phenomenological approach, all of these contributions are absorbed
into an effective $D-K_H$ coupling.  For the non-resonance
contribution to the decay, we will employ the Bauer-Stech-Wirbel (BSW)
model~\cite{Bauer:1987bm}. In our picture all strong phases are
generated by the nearby resonances.  We will focus on the effect of
the closest resonance, $K^*(1945)$.

We will not make any assumptions about the size of $SU(3)$ violation in
the couplings of $K_H$ to $D^0$ and $\D0bar$. We will find from our fit
that the violation is large.  While we propose no explicit mechanism to
account for this, such a scenario need not be unnatural. For instance, one
can imagine  an extension of the models presented in
Refs.~\cite{Buccella:1995nf,Elaaoud:1999pj}. In these models, $K_H$ is
coupled to the charmed mesons via an intermediate state composed of two
ground state mesons, such as $K\pi$,  which is treated in the
factorization approximation, and $SU(3)$ violation is generically small.
However, one should, in general, also consider additional intermediate
states involving higher excitations of kaons and pions. For example, a
contribution of the $K^*_H(1430) \pi$ intermediate state is quite
different in $D^0$ and $\D0bar$ decays even in the factorization approach,
due to the large differences in the values of the relevant form factors
and decay constants. Moreover, phase space effects will differ for
different intermediate states.  Certainly some combination of these
factors could induce an $SU(3)$ violating effective coupling of the
$D$ meson to the $K_H$.  If it is mediated by physical
intermediate states, this coupling generally will be complex; however,
for simplicity we will neglect this additional source of strong
phases.

We begin with a general discussion of the strong phase $\delta$
as a function of the decay amplitudes, which we decompose as
\begin{equation} \label{ampl}
A = A_T + A_R\,e^{i\phi}\,,\qquad B = B_T + B_R\,e^{i\phi}\,.
\end{equation}
Here $A_T$ and $B_T$ are the ``tree'' (nonresonant) contributions,
and $A_R\,e^{i\phi}$ and $B_R\,e^{i\phi}$ are the amplitudes induced by
$K_H$.  The $B$-amplitudes are CF while the $A$ amplitudes are DCS.
We assume here that strong phases arise only from the $K_H(1945)$
propagator in the resonance contribution. Consequently we may take $B_T$
and $A_T$ to be real (in the absence of new physics which might generate
a ``weak'' phase), and the phase $\phi$,
\begin{equation}
   \tan\phi=-{\Gamma_{K_H} m_D\over m_D^2-m_{K_H}^2}=1.23\,.
\label{calphi}
\end{equation}
to be the same in $B_R$ and
$A_R$.  In particular, we assume here that the $D-K_H$ and
$\overline D-K_H$ effective couplings are real.

It is straightforward to express the strong phase difference
$\delta=\delta_B-\delta_A$ in terms of the relative contributions of the
``tree'' processes and the phase of the resonance amplitude:
\begin{equation}
   \tan\delta=\pm\sin\phi\,{\sqrt{|B/B_T|^2-\sin^2\phi}
   -\sqrt{|A/A_T|^2-\sin^2\phi}\over \sin^2\phi+
   \sqrt{|B/B_T|^2-\sin^2\phi}\,\sqrt{|A/A_T|^2-\sin^2\phi}}\,.
\label{tandeamps}
\end{equation}
We parametrize the relative contribution of the resonance amplitude to the
CF decay by
\begin{equation}
   \Zres\equiv{B_R/ B_T}\,.
\label{defres}
\end{equation}
In the $SU(3)$ limit, $A_R/A_T=B_R/B_T$. We parametrize SU(3) violation by
\begin{equation}
   \Zsut\equiv{A_R/A_T\over B_R/B_T}\,,
\label{defsut}
\end{equation}
that is, $A_R/A_T=\Zres\Zsut$. Then we may write
\begin{equation}
   \tan\delta={\sin\phi\Zres(1-\Zsut)\over
   1+\Zres^2\Zsut+\cos\phi\Zres(1+\Zsut)}\,.
\label{tandeZ}
\end{equation}
We clearly see that $\tan\delta$ is not required to be negligible, as is
commonly assumed.  There are, however, three scenarios that
could, in principle, lead to a small strong phase~$\delta$:
\par {(i)} $\sin\phi\ll1$, that is, no nearby resonance.
\par {(ii)} $\Zres\ll1$, that is, negligible resonance contribution.
\par {(iii)} $|\Zsut-1|\ll1$, that is, negligible $SU(3)$ violation.
\par\noindent As is obvious from Eq.~(\ref{calphi}), scenario (i) is not
realized in nature.  Next we will fit our model to experiment and find
that there is no evidence at present that either (ii) or (iii) is realized.

\section{Modeling the strong phase difference}

The starting point for our model of the decay rate is the
effective weak Hamiltonian for $c$ decays,
\begin{eqnarray}
{\cal H}_{eff} &=&
\frac{G_F}{\sqrt{2}} V_{cs} V_{ud}^*
\bigg[ C_1 (\mu)\, \bar s \Gamma_\mu d\, \bar u \Gamma^\mu c +
C_2 (\mu)\, \bar u \Gamma_\mu d\, \bar s \Gamma^\mu c \bigg]
\nonumber \\
&&\mbox{}+ \frac{G_F}{\sqrt{2}} V_{cd} V_{us}^* \bigg[
C_1 (\mu)\, \bar d \Gamma_\mu s\, \bar u \Gamma^\mu c +
C_2 (\mu)\, \bar u \Gamma_\mu s\, \bar d \Gamma^\mu c \bigg],
\end{eqnarray}
where $\Gamma^\mu=\gamma^\mu(1+\gamma^5)$, and
$C_1 (m_D) \simeq -0.514$ and $C_2 (m_D) \simeq 1.270$
in the scheme-independent prescription~\cite{Buccella:1995nf}.  Within the
framework of the BSW model, the nonresonant contributions are given by
\begin{eqnarray} \label{treeamp}
A_T &=& a_1 \frac{G_F}{\sqrt{2}} V_{cd} V_{us}^* f_K
f_+^{D\pi} (m_K^2) m_D^2 \left( 1 -  {m_\pi^2}/{m_D^2}
\right), \nonumber \\
B_T &=& a_1 \frac{G_F}{\sqrt{2}} V_{cs} V_{ud}^* f_\pi
f_+^{DK} (m_\pi^2) m_D^2 \left( 1 -  {m_K^2}/{m_D^2}
\right).
\end{eqnarray}
Here $f_{\pi(K)}$ is the pseudoscalar meson decay
constant, $f_+^{D\pi(K)}$ is the semileptonic
form factor for $D \to \pi(K)$ transitions, and
$a_{1,2}=C_{2,1}+C_{1,2}/N_c$.
In the $SU(3)$ limit, $f_\pi=f_K,~f_+^{D\pi} (m_K^2)=
f_+^{DK} (m_\pi^2)$, and $m_\pi = m_K$. For simplicity, we
have dropped the numerically insignificant contribution
to Eq.~(\ref{treeamp}) from the $f_-$ form factor.

The resonance-mediated amplitudes are given by
\begin{eqnarray} \label{AR}
A_R &=& \frac{g_H}{m_D^2-m_{K_H}^2 + i\Gamma_{K_H}m_D}\, \langle K_H |
H_{eff} | D^0 \rangle\,,
\nonumber \\
B _R &=& \frac{g_H}{m_D^2-m_{K_H}^2 + i\Gamma_{K_H}m_D}\,\langle K_H |
H_{eff} | \D0bar \rangle\,,
\end{eqnarray}
where $K_H=K^*(1950)$.  Here $g_H$ is the coupling of the
resonance $K_H$ to the final state $K^+ \pi^-$, which  may be obtained
from the measured branching ratio of $(52 \pm 14)\%$ for $K_H \to K^+
\pi^-$.  We write the weak matrix elements entering
Eq.~(\ref{AR}) as
\begin{eqnarray}
\langle K_H | H_{eff} | D^0 \rangle &=&
a_2 \frac{G_F}{\sqrt{2}} V_{cd} V_{us}^*
m_D^2 m_{K_H}^2\cdot a\,,
\nonumber \\
\langle K_H | H_{eff} | \D0bar \rangle&=&
a_2 \frac{G_F}{\sqrt{2}} V_{cs} V_{ud}^*
m_D^2 m_{K_H}^2\cdot\bar a\,.
\end{eqnarray}
Here $a$ and $\bar a$ parameterize the effective couplings of
$D^0$ and $\D0bar$ to the resonance $K_H$.  In the $SU(3)$ limit,
$\bar a = a$ and the relative phase between $A$ and $B$
vanishes.  If $SU(3)$ is broken, then
$\delta a=\bar a - a\neq 0$, generating a
relative strong phase between $A$ and $B$.  The values of $\bar a$
and $\delta a$ can be obtained by fitting to the
measured branching ratios ${\cal B}(D^0 \to K^+ \pi^-)$ and ${\cal
B}(\D0bar \to K^+\pi^-)$~\cite{Caso:1998tx}.  Note that we do not include
the usual ``weak scattering'' amplitude of the BSW model.  Instead, this
term is absorbed into our new resonance amplitudes $A_R$ and $B_R$.  This
is the most natural thing to do, since we are essentially proposing a
significant resonance enhancement of this ordinarily small contribution.

For the calculation of $B_T$, we use $f_\pi=132$ MeV,
$f^{D\pi}_+(0)=0.69$, and a pole form to extroplate $f_+$ to the correct
kinematic point~\cite{Bauer:1987bm}.  For the calculation of $A_T$, we
assume $SU(3)$ symmetry in the nonresonant amplitudes, so
$A_T=B_T\tan^2\theta_C$. Then we find $B_R$ and $A_R$ by fitting to the
observed decay rates. We find $|B/B_T|^2=0.76$ and, for the central value
of $R_{\rm exp}=1.58$, we find $|A/A_T|^2=1.24$. The expression
(\ref{tandeamps}) for the phase difference yields
\begin{equation}
   \tan\delta=\pm0.33\quad\Rightarrow\quad\sin\delta=\pm0.31\,.
\end{equation}
In this case, the two solutions for the resonance fraction are
$\Zres=-0.23$ and $-1.03$.  We see that the inclusion of the resonance
amplitudes can have a dramatic impact, even though the couplings $a$ and
$\bar a$ are of the moderate size $|a|\sim |\bar a|\sim {\rm few}\times
10^{-2}$.  The effect is amplified by the small denominators in
Eq.~(\ref{AR}) associated with the presence of the nearby resonance.  The
substantial $SU(3)$ violation in the central value of
$R_{\rm exp}$ is reflected in the fit, with the two solutions yielding
$\Zsut=-0.70$ and $1.38$.

\begin{figure}[t]
\epsfxsize=5in
\hskip0.4in
\epsfbox{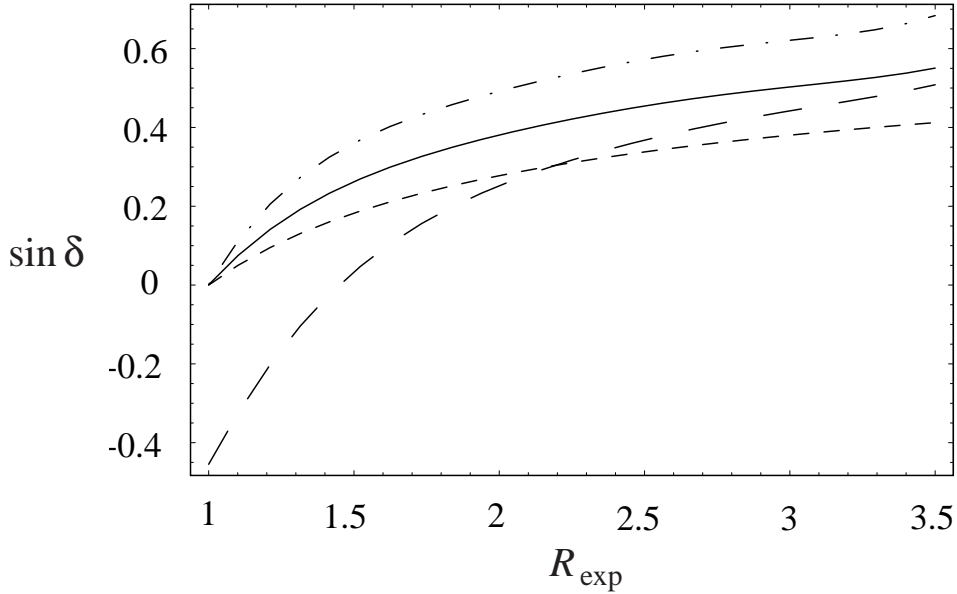}
\caption{The variation of $\sin\delta$ with $R_{\rm
exp}$.  The solid line is our model with $\Gamma(K_H)=200\,$MeV and
$SU(3)$ symmetry in the ``tree'' amplitudes.  The other lines demonstrate
the robustness of the result if various assumptions are relaxed: (i)
$SU(3)$ violation is incorporated into the BSW model (long dashed line);
(ii) we set $\Gamma(K_H)=260\,$ MeV (dashed-dotted line); (iii) the BSW
model is modified so that $|B/B_T|^2=1.0$ (shorted dashed line).  We show
one of the two solutions for $\sin\delta$; the other has the same
magnitude and the opposite sign.}
\label{phase}
\end{figure}

The value of $\sin\delta$ can be rather large in this framework, and it
depends strongly on the amount of $SU(3)$ violation observed in $R_{\rm
exp}$.  It also depends on the width of the resonance $K_H$, through the
value of $\sin\phi$.  The dependence on the details of the BSW model
comes entirely through the value of $B_T$.  Note that  $a_2$
appears only in the combinations $a_2a$ and $a_2\bar a$, which are fit to
experiment; hence our predictions are independent of $a_2$.  Any variation
in the model, such as including or dropping $1/N_c$ terms (which only has
a moderate effect on $a_1$), can be absorbed into a change in $B_T$.
Our model also can be modified to include the effects of $SU(3)$ violation
in the BSW amplitudes $A_T$ and $B_T$.  This is done by taking $f_K =
162~\mbox{MeV} \neq f_\pi$ and $f_+^{DK} (0) = 0.76 \neq f_+^{D\pi}
(0)$~\cite{Bauer:1987bm}.

In Fig.~\ref{phase}, we plot $\sin\delta$ as a function of $R_{\rm exp}$.
We also show how the curves vary if we change the width of $K_H$ to
$260\,$MeV, include $SU(3)$ violation in the BSW amplitudes, or set
$|B/B_T|^2=1$ instead of 0.76.  It is clear from the figure that values of
$\sin\delta$ in the range 0.3 or larger are an entirely generic
consequence of including the resonance $K_H$.  We include in
Fig.~\ref{phase} large values of $R_{\rm exp}$, beyond the range in
Eq.~(\ref{Rexp}), to allow a comparison of our results to previous results
in the literature that were derived for $R_{\rm exp}\sim3$.

%%%%%%%%%%%%%%%%%%%%%%%%%%%%%%%%%%%%
\section{Discussion and Conclusions}

Our model is sufficient to draw a qualitative conclusion about the phase
of $\lambda$, if not a quantitative one. The issue at hand is really
whether, in the CP limit, $\lambda$ can be taken to be approximately real
in the analysis of $D^0-\D0bar$ mixing. If this were so, then the
observation of a substantial CP conserving linear term in $t$ in
$\Gamma(D^0(t)\to K^+\pi^-)$ would indicate the presence of a puzzling new
physics contribution to $\Delta\Gamma$.  Hence one must assess carefully
the robustness of the assumption that $\delta\simeq0$.

While there is a model-indepedent argument that in the $SU(3)$ (and CP)
limit $\lambda$ is real, it is known that this symmetry is not always
well respected in the $D$ system.  In fact, a naive estimate of $SU(3)$
corrections would put them at the level of 30\%, but $\sin\delta$ can only
be ignored in the experimental analysis if it is considerably smaller than
this.  What really must be negligible is the combination
$2\tan\delta\,\Delta M/\Delta\Gamma$.  In the Standard Model, where
$\Delta M/\Delta\Gamma\sim1$, one ought to require at least
$\sin\delta<0.1$.  In new physics scenarios in which $\Delta
M\gg\Delta\Gamma$, the requirement would be substantially stricter.  In
any case, to use $SU(3)$ as an argument for neglecting the phase would
require actually that $SU(3)$ be respected {\it unusually well.}  The
current data on $R_{\rm exp}$ provide no support for such a scenario.

Another argument for a vanishingly small strong phase could be
the absence of a mechanism for generating one.  But given the existence of
nearby resonances, this is not the case for the decay in question. It
could also be that the resonance contributions are very small compared to
the nonresonant ones, or that they respect $SU(3)$. Our fit shows that
this is not necessarily the situation in the $D\to K\pi$ decays. We find
therefore that there is no small parameter which would suppress the
strong phases in these decays.

We have presented a model for the transition $D^0\to K^\pm\pi^\mp$ in
which there is a significant $SU(3)$ violating contribution from an
intermediate excited kaon resonance, and in which the strong phase of
$\lambda$ is large.  While we have designed this model, to
some extent, to yield a significant strong phase difference, we have
neither tuned parameters nor invoked an unnaturally large coupling to
the resonance.  In light of the ease with which this model produces a
reasonably large phase, we would conclude that there is no good reason
to assume $\sin\delta=0$ in any experimental analysis.

Finally, we note that while most of our analysis has been implicitly
carried out in the framework of the Standard Model, our results
are valid even in the presence of new physics. First, it is unlikely
that the strong phase of $\lambda$ is affected by new physics.  There
are certainly extensions of the Standard Model, such as SUSY with
${\cal R}$ parity non-conservation, models with multiple Higgs
doublets, and models with ``exotic'' quarks, which have new tree level
flavor changing interactions that could contribute to the DCS decays.
But the stringent existing constraints from $K^{0}-\overline K{}^{0}$
and $D^{0}-\D0bar$ mixing preclude models in which such
interactions are large enough to compete with standard $W$
exchange~\cite{Bergmann:1999pm}. Second, while the Standard Model
predicts that $\phi$ of Eq. (\ref{lambda}) vanishes to an excellent
approximation, the question of CP violating contributions to
the interference term can be settled experimentally and does not
require any model input. With $\phi\neq0$, we would have
contributions of opposite signs to the interference terms
in $D^0\to K^+\pi^-$ and in $\D0bar\to K^-\pi^+$. In the CP
limit, the respective interference terms are equal. The presence
of CP conserving interference terms can therefore be established
experimentally, and then our comments of how to interpret it in
terms of $\Delta\Gamma$ and $\Delta M$ apply.

%%%%%%%%%%%%%%%%%%%%%%%%%%%%%%%%%%%%%%%%%%%%%%%%%%
\acknowledgments

It is a pleasure to thank Tom Browder, Harry Nelson and Sandip Pakvasa
for useful correspondence. A.F.~and A.P.~are supported in part by the
United States National Science Foundation under Grant No.~PHY-9404057 and
the United States Department of Energy under Outstanding Junior
Investigator Award No.~DE-FG02-94ER40869. A.F.~is a Cottrell Scholar of
the Research Corporation.  Y.N.~is supported by a DOE grant
DE-FG02-90ER40542, by the Ambrose Monell Foundation, by the United
States--Israel Binational Science Foundation (BSF), by the Israel
Science Foundation founded by the Israel Academy of Sciences
and Humanities, and by the Minerva Foundation (Munich).

%%%%%%%%%%%%%%%%%%
\tighten

\end{document}